# Thermodynamics of the SmCo$_5$ compound doped with Fe and Ni: an *ab initio* study

A. Landa[1], P. Söderlind[1], D. Parker[2], D. Åberg[1], V. Lordi[1], A. Perron[1], P. E. A. Turchi[1], R. K. Chouhan[3], D. Paudyal[3], and T. A. Lograsso[3,4]

[1]Lawrence Livermore National Laboratory, Livermore, CA, 94551, USA

[2]Oak Ridge National Laboratory, Oak Ridge, TN, 37831, USA

[3]Ames Laboratory, Ames, IA, 50011, USA

[4]Department of Materials Science and Engineering, Iowa State University, Ames, IA, 50011, USA

**Abstract**

SmCo$_5$ permanent magnets exhibit enormous uniaxial magnetocrystalline anisotropy energy and have a high Curie temperature. However, a low energy product presents a significant drawback in the performance of SmCo$_5$ permanent magnets. In order to increase the energy product in SmCo$_5$, we propose substituting fixed amount of cobalt with iron in a new magnet, SmFe$_3$CoNi, where inclusion of nickel metal makes this magnet thermodynamically stable. We further discuss some basic theoretical magnetic properties of the SmCo$_5$ compound.

**1. Introduction**

Three basic material parameters determine the intrinsic properties of hard magnetic materials: (i) spontaneous (saturation) magnetization, $M_s$, (ii) Curie temperature, $T_c$, and (iii) magnetocrystalline anisotropy energy (MAE), $K_1$ [1]. These three parameters all need to be large for the permanent magnet to be technologically suitable: $T_c \sim \geq 550$ K, $M_s \sim \geq 1$ MA/m, $K_1 \sim \geq 4$ MJ/m$^3$. The desired values of these properties can be achieved by combining transition-metal (TM) with rare-earth-metal (RE) atoms in various intermetallic compounds, where the RE atoms induce a large magnetic anisotropy while the TM atoms provide a large magnetization and high Curie temperature [1, 2].

SmCo$_5$ (in the hexagonal CaCu$_5$-type structure) magnets exhibit enormous uniaxial MAE of $K_1 \sim 17.2$ MJ/m$^3$, substantially higher than that of Nd$_2$Fe$_{14}$B (Neomax) magnets ($K_1 \sim 4.9$ MJ/m$^3$), and have almost twice the Curie temperature ($T_c \sim 1020$ K) compared to Neomax ($T_c \sim 588$ K) [3, 4]. However, Nd$_2$Fe$_{14}$B currently dominates the world market for permanent magnets (~ 62 %) [5, 6], since it possesses the highest energy performance measured by a record energy product *(BH)$_{max}$* of 512 kJ/m$^3$, more than twice as high as the energy product of SmCo$_5$ magnets, *(BH)$_{max}$* of 231 kJ/m$^3$ [3, 4]. Although SmCo$_5$ magnets are more suitable for high temperature applications than Neomax, due to their relatively low energy performance SmCo$_5$ magnets occupy only ~ 3% of the world market [5, 6].

From a cost stand point, it would be beneficial to substitute Co atoms with Fe, because Fe in the Earth's crust is ~ 2000 times more abundant than Co and consequently much cheaper. In addition, Fe is a ferromagnetic metal with a very large magnetization at room temperature (1.76 MA/m [3]). However, the SmFe$_5$ compound is thermodynamically unstable and does not appear in the equilibrium Sm-Fe phase diagram, although the alloy compound Sm(Co$_{1-x}$Fe$_x$)$_5$ with the CaCu$_5$–type structure has been synthesized by the melt-spinning method for $x = 0.0 - 0.3$ [7]. Furthermore, the Curie temperature for Sm(Co$_{1-x}$Fe$_x$)$_5$ alloys was found to increase from ~ 1020 K to ~ 1080 K when increasing $x$ from 0.0 to 0.2 [8], in contrast to Sm$_2$(Co$_{1-x}$Fe$_x$)$_{17}$ alloys which exhibit a monotonic decrease in Curie temperature with increasing Fe content [9].

The fundamental purpose of the present study is to explore the effect of adding Ni to the SmCo$_5$ magnet in order to stabilize Sm(Co$_{1-x}$Fe$_x$)$_5$ alloys containing a sufficient amount of iron to boost the energy product of the Sm(Co-Fe-Ni)$_5$ magnet. We also investigate and report some interesting aspects of the magnetic properties of the SmCo$_5$ compound that were incorrectly described in some previous publications [10-13]. We employ *ab initio* calculations using four complementary techniques: (i) the scalar-relativistic projector-augmented wave (PAW) method implemented in the Vienna *ab initio* simulation package (VASP), (ii) the fully relativistic exact muffin-tin orbital method (FREMTO), (iii) the full-potential linear muffin-tin orbital method (FPLMTO), and (iv) the full-potential linearized augmented plane wave (FPLAPW) method. Both FPLMTO and FPLAPW methods account for all relativistic effects. The usage of these different methods ensures that our results are robust and independent of technical implementations, while taking advantage of the strengths of each method. Pertinent details of the computational methods are described in Section 2. Results of the density-functional calculations of the ground state properties of Sm(Co-Fe-Ni)$_5$ alloys are presented in Section 3. We present results of calculations of the magnetic properties of the SmCo$_5$ compound in Section 4. Lastly, concluding remarks are presented in Section 5.

**2. Computational Details**

Within the Vienna Ab Initio Simulation Package (VASP), the electron-ion interaction is described with the projector augmented wave (PAW) method [14] as implemented by Kresse, Hafner, Furthmüller, and Joubert [15-18]. The PAW potentials treat $5s^2$, $5p^6$, $4f^5$, $5d^1$, $6s^2$ (16 electrons) and $4s^2$, $3d^7$ (9 electrons) as the valence for Sm and Co, respectively. For the electron exchange and correlation energy functional, the generalized gradient approximation (GGA) of Perdew, Burke, and Ernzerhof (PBE) is employed [19]. A planewave energy cutoff of 350 eV is used in calculations. The Brillouin zone is uniformly sampled with a 10x10x11 Monkhorst-Pack [20] grid (*k*-point spacing of 0.15 Å$^{-1}$) together with a Gaussian smearing of 0.2 eV and symmetry turned off. Relaxations of the atomic positions are performed until all forces are below 0.01 eV/Å. Energies are converged to $10^{-6}$ eV. In some cases, non-collinear calculations are performed to evaluate spin-orbit coupling effects (SOC), which are accounted according to the scheme proposed in [21, 22] and non-collinear local moments, with orbital moments computed by projection on the PAW spheres. The magnitudes of the local moments are calculated by projection of the magnetization density onto the PAW spheres, for consistency with the orbital moment calculations. DFT+U on-site potentials are included for Sm 4*f*-states using the approach of Dudarev *et al*. [23] with $U - J = 4.45$ eV [11].

The calculations we refer to as EMTO are performed using the Green's function technique based on the improved screened Korringa-Kohn-Rostoker method, where the one-electron potential is represented by optimized overlapping muffin-tin (OOMT) potential spheres [24, 25]. Inside the potential spheres the potential is spherically symmetric, while it is constant between the spheres. The radius of the potential spheres, the spherical potential inside these spheres, and the constant value in the interstitial region are determined by minimizing (i) the deviation between the exact and overlapping potentials and (ii) the errors caused by the overlap between the spheres. Within the EMTO formalism, the one-electron states are calculated exactly for the OOMT potentials. The outputs of the EMTO calculation include the self-consistent Green's function of the system and the complete, non-spherically symmetric charge density. From this, the total energy is calculated using the full charge-density technique [26]. We treat as valence the *6s*, *5p*, *5d*, and *4f* states for Sm and *4s* and *3d* states for Co and Fe. The corresponding Kohn-Sham orbitals are expanded in terms of *spdf* exact muffin-tin orbitals, *i.e.* we adopt an orbital momentum cutoff $l_{max} = 3$. The EMTO orbitals, in turn, consist of the *spdf* partial waves (solutions of the radial Schrödinger equation for the spherical OOMT potential wells) and the *spdf* screened spherical waves (solutions of the Helmholtz equation for the OOMT muffin-tin zero potential). The completeness of the muffin-tin basis was discussed in detail in Ref. [24]. The generalized gradient approximation (GGA-PBE) [19] is used for the electron exchange and correlation energy. Integration over the Brillouin zone is performed using the special *k*-point technique [27] with 784 *k*-points in the irreducible wedge of the zone (IBZ). The moments of the density of states, needed for the kinetic energy and valence charge density, are calculated by integrating the Green's function over a complex energy contour with 2.0 Ry diameter using a Gaussian integration technique with 30 points on a semi-circle enclosing the occupied states. In the case of the implementation of the FR-EMTO formalism, SOC is included through the four-component Dirac equation [28].

In order to treat compositional disorder, the EMTO method is combined with the coherent potential approximation (CPA) [29, 30]. The ground-state properties of the $Sm(Co_{1-x-y}Fe_xNi_y)_5$ alloys are obtained from EMTO-CPA calculations, including account of the Coulomb screening potential and energy [31-33]. The equilibrium atomic density of the alloy is obtained from a Murnaghan fit to the total energy versus atomic volume curve [34].

The most accurate and fully relativistic calculations are performed using a full-potential approach with no geometrical approximations, where the relativistic effects, including SOC, are accounted through the conventional perturbative scheme [35] that accurately solves the Dirac equation for both the light and heavy lanthanides [36, 37]. For this purpose, we use a version of the FPLMTO [38] in which the "full potential" refers to the use of non-spherical contributions to the electron charge density and potential. This is accomplished by expanding the charge density and potential in cubic harmonics inside non-overlapping muffin-tin spheres and in a Fourier series in the interstitial region. The spin moments are calculated within both these regions while the orbital moment only in the muffin-tin sphere. For Sm metal, we use two energy tails associated with each basis orbital, and these pairs are different for the semi-core states (*5s* and *5p*) and valence states (*6s*, *6p*, *5d*, and *4f*). With this 'double basis' approach, we use a total of six energy tail parameters and a total of 12 basis functions per atom. Spherical harmonic expansions are carried out up to $l_{max}= 6$ for the basis, potential, and charge density. Spin-orbit coupling (SOC) and orbital polarization (OP) are applied only to the *d* and *f* states. The OP has been implemented in the FPLMTO as a parameter-free scheme where an energy term proportional to the square of the orbital moment is added to the total energy functional to account for intra-atomic interactions. Because of the way it is constructed, the orbital polarization often behaves as a magnification of the SOC, resulting in larger and improved orbital moments [37].

For the electron exchange and correlation energy functional, the generalized gradient approximation (GGA-PBE) is used [19].

We have also performed relativistic full-potential linearized augmented plane wave (FP-LAPW) method [39] calculations within the framework of generalized gradient approximation (GGA+PBE) [19], onsite electron correlation (Hubbard parameter), and spin orbit coupling (GGA+U+SOC). Calculations are performed using WIEN2K package [40]. According to Refs. [10-12], we used values of $U = 5.2$ eV and $J = 0.75$ eV to obtain $U - J = 4.45$ eV. The $k$-space integrations have been performed at least with 13×13×15 Brillouin zone mesh that was sufficient for the convergence of total energies and magnetic moments. According to Refs. [11-12], for the SmCo$_5$ compound the sphere radii for Sm and Co were set as 2.115 a.u. and 2.015 a.u., respectively, with force minimization of 3%. The plane-wave cutoff parameters were chosen as $RK_{max} = 9.0$ and $G_{max}=14$ [11-12].

## 2. Ground state properties of Sm(Co-Fe-Ni)$_5$ alloys

The SmCo$_5$ compound crystallizes in the hexagonal CaCu$_5$-type structure with three nonequivalent atomic sites: Sm$_1$-($1a$), Co$_1$-($2c$), Co$_2$-($3g$) (see Figure 1) with 6 atoms per formula unit and also per computational cell.

Table 1 shows the equilibrium formula unit (f.u.) volume, bulk modulus, and the pressure derivative of the bulk modulus for the SmCo$_5$ compound computed using each of the *ab initio* methodologies. The results of the PAW calculations are shown both without ('VASP') and with ('VASP+SOC') inclusion of SOC. Similarly, the results of the EMTO calculations are shown both without ('SREMTO', 'SR' means scalar-relativistic) and with ('FREMTO') inclusion of SOC. The experimental value of the equilibrium f.u. volume is taken from Ref. [41]. One can see that the equilibrium volumes listed in Table 1 agree quite well with the experimental data and that relativistic SOC has only a minor influence. Account for Coulomb correlations in the *f*-shell (FPLAPW) also has only a small influence on the equilibrium volume.

Figure 2 shows the calculated heat of formation of the SmTM$_5$ compounds (TM = Fe, Co, Ni) as a function of the valence 3$d$-electron band occupation. The heat of formation is calculated with respect to the ground-state structures of the pure elements, *i.e.*, $\alpha$-structure of Sm, hexagonal close packed (hcp) Co, body centered cubic (bcc) Fe, and face centered cubic (fcc) Ni. The heat of formation of the SmTM$_5$ compounds drop from about +12 mRy/atom for TM = Fe to about –1 mRy/atom for TM = Co and about –19 mRy/atom for TM = Ni. This tendency could be explained by a shift in the Fermi level due to the gradual filling of the 3$d$-band from 7, to 8, and to 9 electrons for Fe, Co, and Ni, respectively. Both EMTO and FPLMTO methods show almost identical results.

Figure 3 shows the results of calculation of the heat of formation of SmCo$_5$ compounds doped with Fe. The Coherent Potential Approximation (CPA) implemented within the *ab initio* EMTO method allows a gradual substitution of Co atoms by Fe atoms on the Cu-type 3$g$ and 2$c$ sites of the CaCu$_5$–type structure. We see that EMTO-CPA calculations predict a very small region of stability ($x \leq 0.05$) for Sm(Co$_{1-x}$Fe$_x$)$_5$ alloys. FPLMTO calculations for the SmCo$_5$ and SmFe$_5$ pseudo-binary end points give similar results to those given by the EMTO method; a simple straight line interpolation between these points gives a region of stability of $x \leq 0.10$.

Figure 4 shows results of similar calculations for the pseudo-binary Sm(Ni$_{1-x}$Fe$_x$)$_5$ alloys. Due to the significant negative value of the heat of formation of the SmNi$_5$ compound (~ –19 mRy/atom), EMTO-CPA calculations show that approximately half of the Ni atoms in SmNi$_5$ compound can be substituted by Fe atoms. FPLMTO calculations for the SmNi$_5$ and SmFe$_5$ end points show an even larger extent of stable substitution of Ni by Fe. These EMTO-CPA calculations are in excellent agreement with recent SEM/EDS measurements [42], which find that the maximum extension of SmNi$_{5-x}$Fe$_x$ alloys (in the CaCu$_5$-type structure) is about 50 at. % Fe, resulting in formation of the SmFe$_3$Ni$_2$ compound.

Previous neutron diffraction studies of Th(Co$_{1-x}$Fe$_x$)$_5$ alloys (also based on the CaCu$_5$-type structure) [43] show that the larger Fe atoms prefer to occupy the 3$g$-type sites, whereas the smaller Co atoms prefer to occupy the 2$c$-type sites. Figure 5 shows the heat of formation within the CPA formalism of pseudo-binary SmFe$_3$(Ni$_{1-x}$Co$_x$)$_2$ alloys where Fe atoms occupy all 3$g$-type sites and the occupation of the 2$c$-type sites gradually changes from pure Ni (the SmFe$_3$Ni$_2$ compound) to pure Co (the SmFe$_3$Co$_2$ compound). Present calculations show that SmFe$_3$(Ni$_{1-x}$Co$_x$)$_2$ alloys could remain stable until approximately half of Ni atoms are substituted by Co atoms. In the next section, we will discuss why it is important to reach maximum solubility (~ 50 at. %) of Fe atoms in SmNi$_{5-x}$Fe$_x$ alloys (in other words to create SmFe$_3$Ni$_2$ compound with Fe atoms occupying the 3$g$-type sites) before starting substitution of Ni atoms by Co. We will also discuss the possibility to achieve high axial MAE for energetically stable SmFe$_3$(Ni$_{1-x}$Co$_x$)$_2$ alloys using more abundant and cheap Fe and Ni than Co, and simultaneously reaching a higher magnetic energy product than that of the initial SmCo$_5$ prototype.

## 4. Magnetic properties of the SmCo$_5$ compound.

Table 2 presents previously reported results [10-12] of the site-projected spin, $m^{(s)}$, and orbital, $m^{(o)}$, moments of the SmCo$_5$ compound calculated by FPLAPW. The calculated total moment $m^{(tot)} = 9.90$ $\mu_B$/f.u. significantly exceeds reported experimental values of 7.80 $\mu_B$/f.u. [44] and 8.9 $\mu_B$/f.u. [45]. Also, the calculated MAE, $K_1 = 40.79$ MJ/m$^3$, substantially exceeds the experimental value $K_1 = 17.2$ MJ/m$^3$ [3]. We repeated these calculations using both FPLAPW and VASP-PAW methods within the LDA+U formalism, using identical $U - J = 4.45$ eV as in Refs. [10-12]. The results for both (001) and (100) orientations are presented in Tables 3 and 4 (FPLAPW) and Tables 5 and 6 (VASP). As in the case of the original FPLAPW calculations [10-12], we assume parallel spin alignment of Sm and Co atoms. Our calculated MAE are $K_1 = 35.96$ MJ/m$^3$ for FPLAPW and $K_1 = 43.43$ MJ/m$^3$ for VASP-PAW, both significantly higher than the experimental value $K_1 = 17.2$ MJ/m$^3$ [3], consistent with the prior reported results.

Although both FPLAPW and VASP calculations qualitatively predict the correct axial anisotropy of the SmCo$_5$ compound, we think that these MAE results are "artificial" because of the incorrect parallel alignment between the spins of Sm and Co atoms. It is well established that spins of RE and TM atoms always align antiparallel in RE-TM compounds [46, 47]. In order to confirm this fact, we performed calculations of the total energy of the SmCo$_5$ compound as a function of the atomic volume for both parallel and antiparallel Sm and Co spin alignments. Results of these calculations, performed by both FPLMTO and FPLAPW methods (with $U = J = 0$), are shown in Figures 6 and 7, respectively. The results show that the parallel spin alignment, adopted in Refs. [10-12], represents a *metastable* state, which lies about 7 mRy/atom above the *ground* state with antiparallel spin orientation. FPLAPW LDA+U calculations, performed with $U - J = 4.45$ eV, result in a similar conclusion (see Figure 8): within this level of theory, the *metastable* parallel spins state lies ~ 4 mRy/atom above the antiparallel spins *ground* state.

Tables 7 – 12 present the results of the site-projected spin, $m^{(s)}$, and orbital, $m^{(o)}$, moments of the SmCo$_5$ compound calculated by the different methods assuming antiparallel Sm and Co spin alignment. In the case of VASP+SOC, FREMTO, and FPLMTO calculations ($U = J = 0$), as shown in Tables 7, 8, and 9, respectively, the total moment is 4.18 $\mu_B$/f.u., 4.60 $\mu_B$/f.u., and 4.04 $\mu_B$/f.u., respectively, which are all less than half of the reported experimental values of 7.80 $\mu_B$/f.u. [44] or 8.9 $\mu_B$/f.u. [45]. Accounting for Coulomb correlations in the *f*-shell within the FPLAPW method ($U - J = 4.45$ eV) produces a higher value of the total magnetic moment of

5.56 $\mu_B$/f.u. (Table 11), closer but still somewhat below the experimental values. Significant increase of the total magnetic moment to $m^{(tot)}$ = 6.09 $\mu_B$/f.u. is achieved by incorporating orbital polarization (OP) [48, 49] in the FPLMTO ($U = J = 0$) method (see Table 10).

FPLAPW ($U - J$ = 4.45 eV) calculations (Table 11) predict the MAE to be negative (in-plane orientation) with $K_1 \approx$ –0.73 mRy/f.u. $\approx$ –18.75 MJ/m$^3$. To understand these failures to accurately describe the total moment of the SmCo$_5$ compound (see Table 7 – Table 11) and its MAE (in-plane orientation), even when accounting for the energetically stable antiparallel Sm and Co spin orientation, we need to examine several well-established experimental facts. Polarized nuclear magnetic resonance (NMR) studies showed that the large anisotropy of the SmCo$_5$ compound originates from a large orbital contribution from Co$_1$(*2c*) sites that are located in the same plane as Sm$_1$(*1a*) atoms [50-52]. As a result, Co$_1$(*2c*) atoms have a large positive anisotropy contribution while Co$_2$(*3g*) atoms have a small negative anisotropy contribution. This experimental observation has been confirmed by several calculations for the related YCo$_5$ compound [53, 54]. However, the calculations presented above in Tables 7 – 11 predict the opposite tendency: the orbital moment of Co$_2$(*3g*) atoms is equal to (Table 7) or larger than (Tables 8 –11) the orbital moment of Co$_1$(*2c*) atoms, which *a priori* cannot produce an axial (positive) MAE for the SmCo$_5$ compound.

In a previous work [55], the authors claim that neither LDA nor LDA+U are sufficient for describing the 4*f* states of the SmCo$_5$ compound and, in order to describe the crystal field (CF) effects of the localized 4*f* shell, a more sophisticated LDA+DMFT (dynamical mean-field theory) method is necessary. According to these calculations, reproduced in Table 12, the orbital moments of Co$_1$(*2c*) are larger than the orbital moments of Co$_2$(*3g*) and also the calculated total moment, $m^{(tot)}$ = 8.02 $\mu_B$/f.u., is significantly larger than that obtained above, even with opposite Sm and Co spin orientations (see Tables 7 – 11), although still a bit lower than the experimental value of $m^{(tot)}$ = 8.9 $\mu_B$/f.u. [45]. We can explain the remaining discrepancy by the fact that, as seen in Table 12, the absolute value of the spin moment of Sm atoms is still larger than the absolute value of the orbital moment of Sm atoms. As was mentioned in [46, 47], for the light rare earths including Sm, ($J = L - S$), the total moments of RE (Sm) and TM (Co) atoms show ferromagnetic behavior (c.f., Figure 4 in Ref. [47]). Thus, we expect that further improvement in the description of the CF effects of the localized 4*f* electrons of Sm should result in an increase of the absolute value of the orbital moment of Sm, so that (i) the value of the total moment of the

SmCo$_5$ compound will be increased and (ii) the ferromagnetic behavior of the total moments of Sm and Co atoms will be correctly described in accord with experimental observation.

5.**Conclusions**

In this work, we suggest a means to realize a permanent magnet exhibiting both high MAE and Curie temperature comparable to the SmCo$_5$ compound but with a significantly higher energy product. The basic idea is to start from the SmNi$_5$ compound (in the CaCu$_5$-type structure) and dissolve a maximum amount of iron metal (~ 50 at. %) to form the stable SmFe$_3$Ni$_2$ compound in the same structure modification, where iron atoms predominantly occupy 3$g$ sites (in the ideal situation iron atoms should occupy all 3$g$ positions). Subsequent gradual alloying of the SmFe$_3$Ni$_2$ compound with Co, while keeping the amount of Sm and Fe constant, should allow to reach the maximum solubility of cobalt metal, which will substitute Ni atoms on the *2c* sites. Our calculations show that approximately half of the Ni atoms can be replaced by Co atoms.

We believe that even more accurate calculations in the future will show further a predominant influence of the TM (2$c$) sites on the MAE of Sm(TM)$_5$ magnets. Indeed, the application of LDA+DMFT approach to the calculation of MAE of YCo$_5$ magnets by Zhu *et al.* [56] showed that axial (positive) MAE could be obtained only if orbital moments on Co$_1$(2$c$) atoms are always larger than the orbital moments of Co$_2$(*3g*) atoms. The data from Ref. [55] in Table 12 demonstrated a good chance for correct calculation of MAE of SmCo$_5$ magnet by applying LDA+DMFT.

**Acknowledgements**

A. L. thanks A. Ruban, O. Peil, and L. Vitos for technical support. This research is supported by the Critical Materials Institute, an Energy Innovation Hub funded by the US Department of Energy, Office of Energy Efficiency and Renewable Energy, Advanced Manufacturing Office. Part of the work is performed under the auspices of the US Department of Energy by the Lawrence Livermore National Laboratory under Contract No. DE-AC52-07NA27344 and Oak Ridge National Laboratory under Contract No. DE-AC05-00OR22725. Part of the work is performed at the Ames Laboratory, which is operated for the US Department of Energy by Iowa State University under contract DE-AC02-07CH11358.

**References**

1. K. H. J. Buschow, Rep. Prog. Phys. **54**, 1123 (1991).
2. K. H. J. Buschow and F. R. de Boer, Physics of Magnetism and Magnetic Materials, (Kluwer Academic/Plenum Publishers, New York, 2004).
3. J. M. D. Coey, IEEE Trans. Magn. **47**, 4671 (2011).
4. J. M. D. Coey, Scr. Mater. **67**, 524 (2012).
5. O. Gutfleisch, M. A. Willard, E. Brück, C. H. Chen, S. G. Sankar, and J. P. Liu, Adv. Mater. **23**, 821 (2011).
6. M. J. Kramer, R. W. McCallum, I. A. Anderson, and S. Constantinides, JOM **64**, 752 (2012).
7. T. Miyazaki, M. Takahashi, X. Yang, H. Saito, and Takahashi, J. Magn. Magn. Mater. **75**, 123 (1988).
8. T. Miyazaki, M. Takahashi, X. Yang. H. Saito, and M. Takahashi, J. Appl. Phys. **64**, 5974 (1988).
9. A. E. Ray and K. J. Strnat, IEEE Trans. Magn. **8**, 516 (1972).
10. P. Larson and I. I. Mazin, J. Appl. Phys. **93**, 6888 (2003).
11. P. Larson, I. I. Mazin, and D. A. Papaconstantopoulos, Phys Rev. B **67**, 214405 (2003).
12. P. Larson, I. I. Mazin, and D. A. Papaconstantopoulos, Phys Rev. B **69**, 134408 (2004).
13. W. Cheng, S. Zhao, X. Cheng, and X. Miao, J. Supercond. Nov. Magn., **25**, 1947 (2012).
14. P. E. Blochl, Phys. Rev. B **50**, 17953 (1994).
15. G. Kresse and J. Hafner, Phys. Rev. B **47**, 558 (1993).
16. G. Kresse and J. Furhmüller, Phys. Rev. B **54**, 11169 (1996).
17. G. Kresse and J. Furhmüller, Comput. Mater. Sci. **6**, 15 (1996).
18. G. Kresse and D. Joubert, Phys. Rev. B **59**, 1758 (1999).
19. J. P. Perdew, K. Burke, M. Ernzerhof, Phys. Rev. Lett. **77**, 3865 (1996).
20. H. J. Monkhorst and J. D. Pack, Phys. Rev. B **13**, 5188 (1976).
21. L. Kleiman, Phys. Rev. B, **21**, 2630 (1980).
22. A. H. Macdonald, W. E. Pickett, and D. D. Koelling, J. Phys. C: Solid St. Phys., **13**, 2675 (1980).
23. S. L. Dudarev, G. A. Botton, S. Y. Savrasov, C. J. Humphreys, and A. P. Sutton, Phys. Rev. B **57**, 1505 (1998).
24. L. Vitos, Phys. Rev. B **64**, 014107 (2001).

25. L. Vitos, Computational Quantum Mechanics for Materials Engineers: The EMTO Method and Application, (Springer-Verlag, London, 2007).

26. J. Kollar, L. Vitos, H. L. Skriver, in: H. Dreyssé (Ed.), Electronic Structure and Physical Properties of Solids: The Uses of the LMTO Method, Lecture Notes in Physics, (Springer-Verlag, Berlin, 2000), pp. 85-113.

27. D. J. Chadi and M. L. Cohen, Phys. Rev. B **8**, 5747 (1973); S. Froyen, Phys. Rev. B **39**, 3168 (1989).

28. L. V. Pourovskii, A. V. Ruban, L. Vitos, H. Ebert, B. Johansson, and I. A. Abrikosov, Phys. Rev. B **71**, 094415 (2005).

29. J. S. Faulkner, Prog. Mater. Sci. **27**, 1 (1982).

30. L. Vitos, I. A. Abrikosov, and B. Johansson, Phys. Rev. Lett. **87**, 156401 (2001).

31. A. V. Ruban and H. L. Skriver, Phys. Rev. B **66**, 024201 (2002).

32. A. V. Ruban, S. I. Simak, P. A. Korzhavyi, H. L. Skriver, Phys. Rev. B **66**, 024202 (2002).

33. A. V. Ruban, S. I. Simak, S. Shallcross, and H. L. Skriver, Phys. Rev. B **67**, 214302 (2003).

34. F. D. Murnaghan, Proc. Natl. Acad. Sci. U.S.A. **30**, 244 (1944).

35. O. K. Andersen, Phys. Rev. B **12**, 3060 (1975).

36. P. Söderlind, Phys. Rev. B **65**, 115105 (2002).

37. P. Söderlind, P. E. A. Turchi, A. Landa, and V. Lordi, J. Phys.: Condens. Matter **26**, 416001 (2014).

38. J. M. Wills, M. Alouani, P. Andersson, A. Delin, O. Eriksson, and O. Grechnev, Full-Potential Electronic Structure Method, (Springer-Verlag, Berlin, 2010).

39. D. J. Singh, Planewaves, Pseudopotentials, and the LAPW Method, (Kluwer Academic, Boston, 1994).

40. P. Blaha, K. Schwarz, G.K.H. Madsen, D. Kvasnicka, J. Luitz, WIEN2k, An Augmented Plane Wave Plus Local Orbitals Program for Calculating CrystalProperties, Vienna University of Technology, Austria, 2013. <http://www.wien2k.at>.

41. A. V. Andreev and S. M. Zadvorkin, Physica B **172**, 517 (1991).

42. K. Nouri, M. Jemmali, S. Walha, K. Zehani, L. Bessais, and A. Ben Salah, J. Alloys Compd. **661**, 508 (2016).

43. J. Laforest and J. S. Shah, IEEE Trans. Magn. **9**, 217 (1973).

44. M. D. Coey, Magnetism and Magnetic Materials, (University Press, Cambridge 2010).
45. T. -S. Zhao, H. -M. Jin, G.-H. Gua, X. -F. Han, and H. Chen, Phys. Rev. B **43**, 8593 (1991).
46. K. Kumar, J. Appl. Phys. **63**, R13 (1988).
47. M. S. S. Brooks, O. Eriksson, and B. Johansson, J. Phys.: Condens. Matter **1**, 5861 (1989).
48. O. Eriksson, B. Johansson, and M. S. S. Brooks, J. Phys.: Condens. Matter **1**, 4005 (1989).
49. O. Eriksson, M. S. S. Brooks, and B. Johansson, Phys. Rev. B **41**, 9087 (1990).
50. J. Déportes, D. Givord, J. Scweizer, and F. Tasset, IEEE Trans. Magn. **6**, 1000 (1976).
51. R. L. Streever, Phys. Lett. A **65**, 360 (1978).
52. R. L. Streever, Phys Rev. B **19**, 2704 (1979).
53. L. Nordström, M. S. S. Brooks, and B. Johansson, J. Phys.: Condens. Matter 4, 3261 (1992).
54. L. Steinbeck, M. Richter, and H. Eschrig, Phys. Rev. B **63**, 184431 (2001).
55. O. Grånäs, I. Di Marco, P. Thunström, L. Nordström, O. Eriksson, T. Björkman, and J. M. Wills, Comp. Mat. Sci. **55**, 295 (2012).
56. J-X. Zhu, M. Janoscheck, R. Rosenberg, F. Ronning, J. D. Thomson, M. Torrez, E. D. Bauer, and C. D. Bastia, Phys. Rev. X **4**, 021027 (2014).
57. N. Rosner, D. Koudela, U. Schwarz, A. Handstein, M. Hafland, I. Opahle, K. Koepernik, M. D. Kuz'min, K,-H. Müller, J. A. Mydosh, and M. Richter, Nature Physics **2**, 469 (2006).

**Figures.**

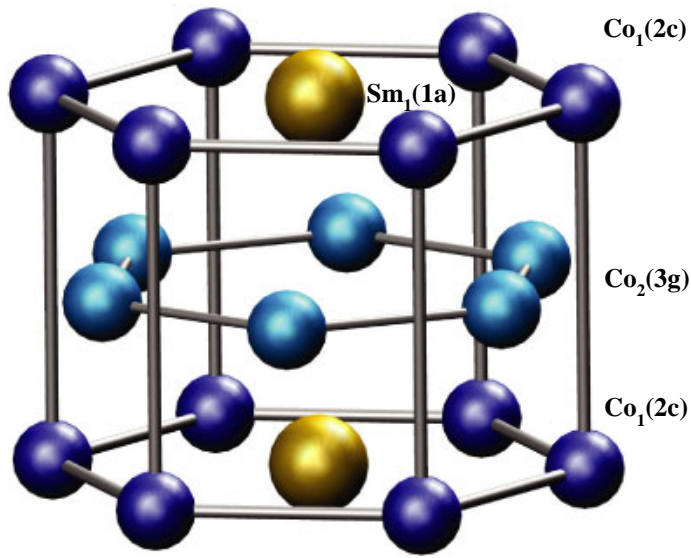

Figure 1. Crystal structure of SmCo$_5$ compound [57].

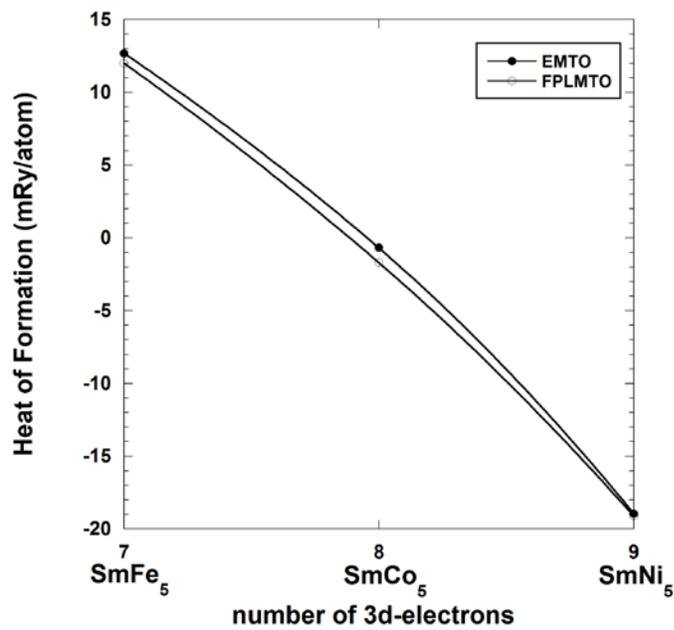

Figure 2. The heat of formation of the SmTM$_5$ compound (TM = Fe, Co, Ni) as a functions of the valence 3$d$-electron band occupation.

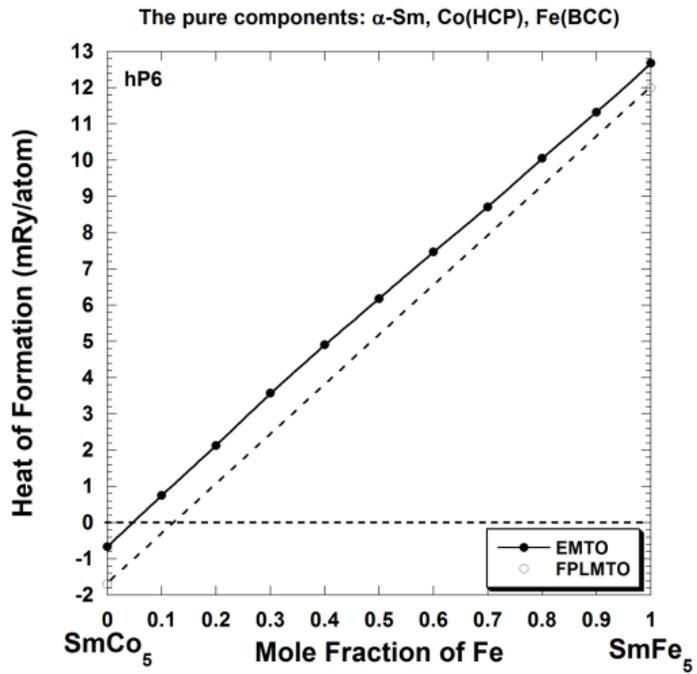

Figure 3. The heat of formation of pseudo-binary Sm(Co$_{1-x}$Fe$_x$)$_5$ alloys.

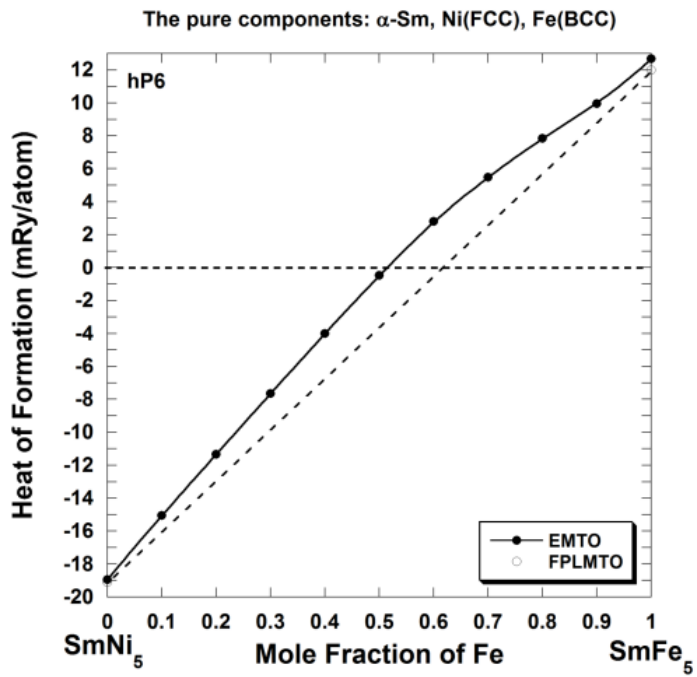

Figure 4. The heat of formation of pseudo-binary Sm(Co$_{1-x}$Ni$_x$)$_5$ alloys.

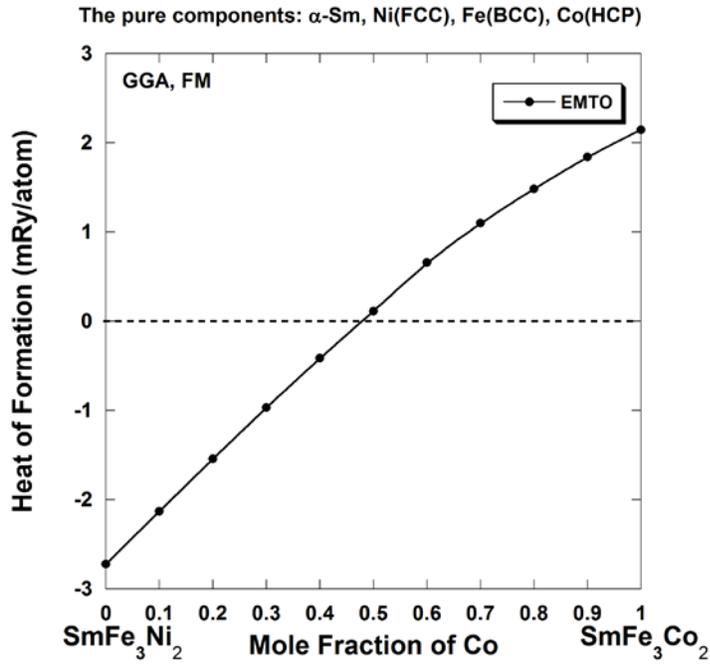

Figure 5. The heat of formation of pseudo-binary $SmFe_3(Ni_{1-x}Co_x)_2$ alloys.

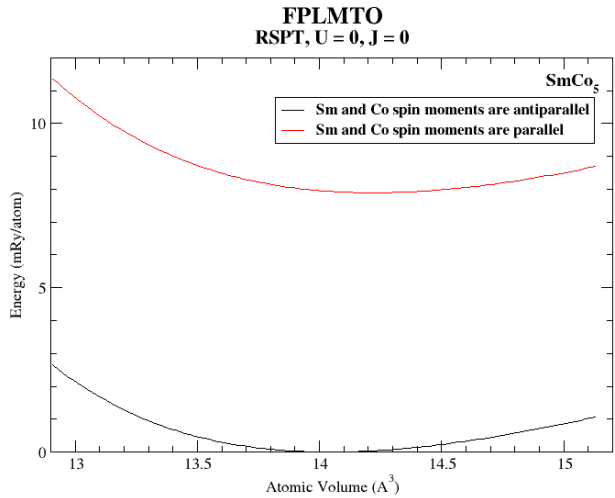

Figure 6. The total energy of the $SmCo_5$ compound as a function of the atomic volume. FPLMTO calculations.

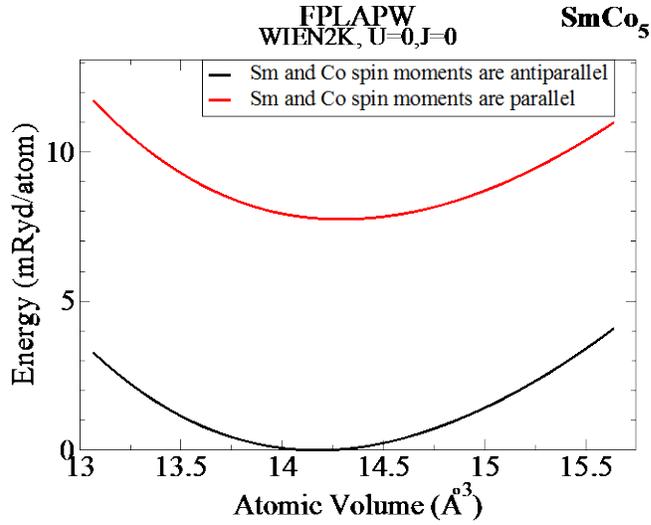

Figure 7. The total energy of the SmCo$_5$ compound as a function of the atomic volume. FPLAPW calculations..

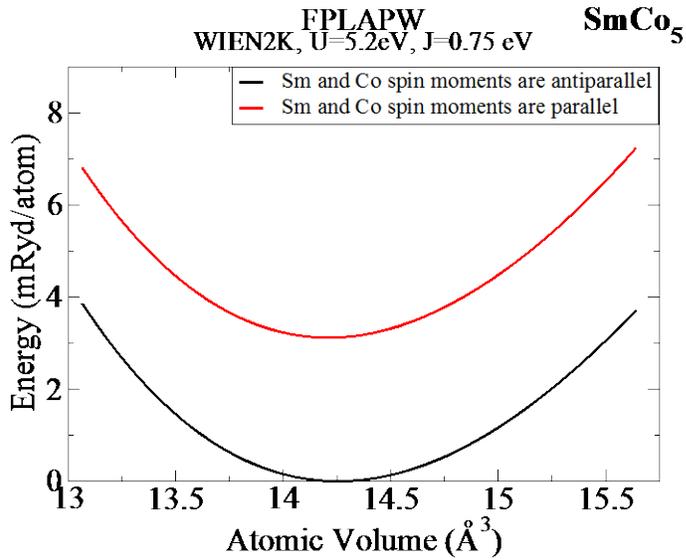

Figure 8. The total energy of the SmCo$_5$ compound as a function of the atomic volume. FPLAPW calculations.

**Tables.**

Table 1. Equilibrium formula unit volume, bulk modulus, its pressure derivative of the SmCo$_5$ compound as functions of the *ab initio* methodology.

| Property | VASP | VASP+SOC | SREMTO | FREMTO | FPLMTO | FPLAPW U=J=0 eV | FPLAPW U=5.20 eV J=0.75 eV | Expt. [41] |
|---|---|---|---|---|---|---|---|---|
| Unit cell volume (Å$^3$) | 84.20 | 84.25 | 85.86 | 85.89 | 84.41 | 85.04 | 85.58 | 85.74 |
| Bulk modulus (GPa) | 134.5 | 134.1 | 133.0 | 130.4 | 123 | 141.1 | 143.1 | |
| Bulk modulus pressure derivative | 4.71 | 4.72 | 3.71 | 3.41 | 3.00 | 4.84 | 4.77 | |

Table 2. Site-projected spin, $m^{(s)}$, and orbital, $m^{(o)}$, moments for the SmCo$_5$ compound: FPLAPW calculations [10-12]. $U - J = 4.45$ eV. $m^{(tot)} = 9.90$ $\mu_B$/f.u. <u>Notice that the parallel spin alignment of Sm and Co atoms represents the metastable state</u>.

| Component: | Sm$_1$(*1a*) | Co$_1$(*2c*) | Co$_2$(*3g*) |
|---|---|---|---|
| $m^{(s)}$ ($\mu_B$) | + 4.46 | + 1.53 | + 1.56 |
| $m^{(o)}$ ($\mu_B$) | - 2.80 | + 0.10 | + 0.10 |

Table 3. Site-projected spin, $m^{(s)}$, and orbital, $m^{(o)}$, moments for the SmCo$_5$ compound: FPLAPW calculations for (001) orientation. $U - J = 4.45$ eV. $m^{(tot)} = 10.57$ $\mu_B$/f.u. Notice that the parallel spin alignment of Sm and Co atoms represents the metastable state.

| Component: | Sm$_1$(1a) | Co$_1$(2c) | Co$_2$(3g) | Interstitial |
|---|---|---|---|---|
| $m^{(s)}$ ($\mu_B$) | + 4.880 | + 1.534 | + 1.555 | - 0.404 |
| $m^{(o)}$ ($\mu_B$) | - 2.277 | + 0.146 | + 0.127 | N/A |

Table 4. Site-projected spin, $m^{(s)}$, and orbital, $m^{(o)}$, moments for the SmCo$_5$ compound: FPLAPW calculations for (100) orientation. $U - J = 4.45$ eV. $m^{(tot)} = 10.56$ $\mu_B$/f.u. Notice that the parallel spin alignment of Sm and Co atoms represents the metastable state.

| Component: | Sm$_1$(1a) | Co$_1$(2c) | Co$_2$(3g) | Interstitial |
|---|---|---|---|---|
| $m^{(s)}$ ($\mu_B$) | + 4.886 | + 1.541 | + 1.546 | - 0.404 |
| $m^{(o)}$ ($\mu_B$) | - 2.163 | + 0.103 | + 0.104 | N/A |

Table 5. Site-projected spin, $m^{(s)}$, and orbital, $m^{(o)}$, moments for the SmCo$_5$ compound: VASP+SOC calculations for (001) orientation. $U - J = 4.45$ eV. $m^{(tot)} = 12.41$ $\mu_B$/f.u. Notice that the parallel spin alignment of Sm and Co atoms represents the metastable state.

| Component: | Sm$_1$(1a) | Co$_1$(2c) | Co$_2$(3g) | Interstitial |
|---|---|---|---|---|
| $m^{(s)}$ ($\mu_B$) | + 5.055 | + 1.509 | + 1.495 | - 0.147 |
| $m^{(o)}$ ($\mu_B$) | - 3.154 | + 0.145 | + 0.121 | N/A |

Table 6. Site-projected spin, $m^{(s)}$, and orbital, $m^{(o)}$, moments for the SmCo$_5$ compound: VASP+SOC calculations for (100) orientation. $U - J = 4.45$ eV. $m^{(tot)} = 12.31$ $\mu_B$/f.u. <u>Notice that the parallel spin alignment of Sm and Co atoms represents the metastable state</u>.

| Component: | Sm$_1$(1a) | Co$_1$(2c) | Co$_2$(3g) | Interstitial |
|---|---|---|---|---|
| $m^{(s)}$ ($\mu_B$) | + 5.000 | + 1.502 | + 1.45 | - 0.150 |
| $m^{(o)}$ ($\mu_B$) | - 2.895 | + 0.102 | + 0.128 | N/A |

Table 7. Site-projected spin, $m^{(s)}$, and orbital, $m^{(o)}$, moments for the SmCo$_5$ compound: VASP+SOC calculations. $U = J = 0$. $m^{(tot)} = 4.18$ $\mu_B$/f.u.

| Component: | Sm$_1$(1a) | Co$_1$(2c) | Co$_2$(3g) | Interstitial |
|---|---|---|---|---|
| $m^{(s)}$ ($\mu_B$) | - 5.97 | + 1.53 | + 1.54 | + 0.47 |
| $m^{(o)}$ ($\mu_B$) | + 1.50 | + 0.10 | + 0.10 | N/A |

Table 8. Site-projected spin, $m^{(s)}$, and orbital, $m^{(o)}$, magnetic moments for the SmCo$_5$ compound: FREMTO calculations. $U = J = 0$. $m^{(tot)} = 4.60$ $\mu_B$/f.u.

| Component: | Sm$_1$(1a) | Co$_1$(2c) | Co$_2$(3g) |
|---|---|---|---|
| $m^{(s)}$ ($\mu_B$) | - 5.74 | + 1.58 | + 1.60 |
| $m^{(o)}$ ($\mu_B$) | + 1.88 | + 0.11 | + 0.12 |

Table 9. Site-projected spin, $m^{(s)}$, and orbital, $m^{(o)}$, magnetic moments for the $SmCo_5$ compound: FPLMTO calculations [55]. $U = J = 0$. $m^{(tot)} = 4.04$ $\mu_B/f.u.$

| Component: | $Sm_1(1a)$ | $Co_1(2c)$ | $Co_2(3g)$ | Interstitial |
|---|---|---|---|---|
| $m^{(s)}$ ($\mu_B$) | - 5.48 | + 1.58 | + 1.55 | - 0.52 |
| $m^{(o)}$ ($\mu_B$) | + 1.81 | + 0.06 | + 0.10 | N/A |

Table 10. Site-projected spin, $m^{(s)}$, and orbital, $m^{(o)}$, magnetic moments for the $SmCo_5$ compound: FPLMTO+OP calculations. $U = J = 0$. $m^{(tot)} = 6.09$ $\mu_B/f.u.$

| Component: | $Sm_1(1a)$ | $Co_1(2c)$ | $Co_2(3g)$ | Interstitial |
|---|---|---|---|---|
| $m^{(s)}$ ($\mu_B$) | - 4.80 | + 1.53 | + 1.56 | - 0.84 |
| $m^{(o)}$ ($\mu_B$) | + 3.30 | + 0.12 | + 0.15 | N/A |

Table 11. Site-projected spin, $m^{(s)}$, and orbital, $m^{(o)}$, magnetic moments for the $SmCo_5$ compound: FPLAPW calculations. $U - J = 4.45$ eV. $m^{(tot)} = 5.56$ $\mu_B/f.u.$

| Component: | $Sm_1(1a)$ | $Co_1(2c)$ | $Co_2(3g)$ | Interstitial |
|---|---|---|---|---|
| $m^{(s)}$ ($\mu_B$) | - 5.339 | + 1.586 | + 1.612 | - 0.721 |
| $m^{(o)}$ ($\mu_B$) | + 3.101 | + 0.073 | + 0.123 | N/A |

Table 12. Site-projected spin, $m^{(s)}$, and orbital, $m^{(o)}$, magnetic moments for the $SmCo_5$ compound: LDA+DMFT calculations [55]. $U - J = 7.03$ eV. $m^{(tot)} = 8.02$ $\mu_B$/f.u.

| Component: | $Sm_1(1a)$ | $Co_1(2c)$ | $Co_2(3g)$ | Interstitial |
|---|---|---|---|---|
| $m^{(s)}$ ($\mu_B$) | - 3.47 | + 1.54 | + 1.52 | - 0.39 |
| $m^{(o)}$ ($\mu_B$) | + 3.26 | + 0.22 | + 0.18 | N/A |